# Capturing Topology in Graph Pattern Matching

Shuai Ma[1]　　Yang Cao[1]　　Wenfei Fan[2]　　Jinpeng Huai[1]　　Tianyu Wo[1]
[1]NLSDE Lab, Beihang University　　　[2]University of Edinburgh
{mashuai@act., caoyang@act., huaijp@, woty@act.}buaa.edu.cn　　　wenfei@inf.ed.ac.uk

## Abstract

Graph pattern matching is often defined in terms of subgraph isomorphism, an NP-complete problem. To lower its complexity, various extensions of graph simulation have been considered instead. These extensions allow pattern matching to be conducted in cubic-time. However, they fall short of capturing the topology of data graphs, *i.e.*, graphs may have a structure drastically different from pattern graphs they match, and the matches found are often too large to understand and analyze. To rectify these problems, this paper proposes a notion of *strong simulation*, a revision of graph simulation, for graph pattern matching. (1) We identify a set of criteria for preserving the topology of graphs matched. We show that strong simulation preserves the topology of data graphs and finds a bounded number of matches. (2) We show that strong simulation retains the same complexity as earlier extensions of simulation, by providing a cubic-time algorithm for computing strong simulation. (3) We present the locality property of strong simulation, which allows us to effectively conduct pattern matching on distributed graphs. (4) We experimentally verify the effectiveness and efficiency of these algorithms, using real-life data and synthetic data.

## 1. Introduction

Graph pattern matching is being increasingly used in a number of applications, *e.g.*, software plagiarism detection, biology, social networks and intelligence analysis [27, 32, 33, 35]. Given a pattern graph $Q$ and a data graph $G$, it is to find all subgraphs of $G$ that match $Q$. Here matching is typically defined in terms of *subgraph isomorphism* (see, *e.g.*, [4, 20] for surveys): a subgraph $G_s$ of $G$ *matches* $Q$ if there exists a *bijective function* $f$ from the nodes of $Q$ to the nodes in $G_s$ such that (1) for each pattern node $u$ in $Q$, $u$ and $f(u)$ have the same label, and (2) there exists an edge $(u, u')$ in $Q$ if and only if $(f(u), f(u'))$ is an edge in $G_s$.

However, subgraph isomorphism is an NP-complete problem [34]. Moreover, there are possibly exponential many subgraphs in $G$ that match $Q$. In addition, as observed in [6, 19], it is often too restrictive to catch sensible matches, as it requires matches to have exactly the same topology as a pattern graph. These hinder its applicability in emerging applications such as social networks and crime detection.

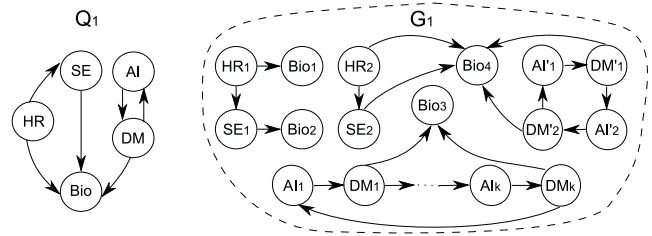

**Figure 1: Social matching: query and data graphs**

To reduce the complexity, *graph simulation* [29] has been adopted for pattern matching. A graph $G$ matches a pattern $Q$ via *graph simulation* if there exists a binary *relation* $S \subseteq V_Q \times V$, where $V_Q$ and $V$ are the set of nodes in $Q$ and $G$, respectively, such that (1) for each $(u, v) \in S$, $u$ and $v$ have the same label; and (2) for each node $u$ in $Q$, there exists $v$ in $G$ such that (a) $(u, v) \in S$, and (b) for each edge $(u, u')$ in $Q$, there exists an edge $(v, v')$ in $G$ such that $(u', v') \in S$. Graph simulation can be determined in quadratic time [24]. Recently this notion has been extended by mapping edges in $Q$ to (bounded) paths in $G$ [19, 18], with a cubic-time complexity, to identify matches in, *e.g.*, social networks.

Nevertheless, the low complexity comes with a price: (1) simulation and its extensions [19, 18] do not preserve the topology of data graphs; in other words, they may match a graph $G$ and a pattern $Q$ with drastically different structures. (2) The match relation $S$ is often too large to understand and analyze, as illustrated below.

**Example 1:** Consider a real-life example taken from [31]. A headhunter wants to find a biologist (Bio) to help a group of software engineers (**SE**s) analyze genetic data. To do this, she uses an expertise recommendation network $G_1$, as depicted in Fig. 1. In $G_1$, a node denotes a person labeled with expertise, an edge indicates recommendation, *e.g.*, $HR_1$ recommends $Bio_1$, and there is an edge from each $DM_i$ to $Bio_3$. The biologist Bio needed is specified with a pattern graph $Q_1$, also shown in Fig. 1. Intuitively, the Bio has to be recommended by: (a) an HR person; (b) an SE, *i.e.*, the Bio has experience working with **SE**s; and (c) a data mining specialist (DM), as data mining techniques are required for the job. Moreover, (d) the SE is also recommended by an HR person, and (e) there is an artificial intelligence expert (AI) who recommends the DM and is recommended by a DM.

When subgraph isomorphism is used, no match can be found, *i.e.*, no subgraph of $G_1$ is isomorphic to $Q_1$. In other words, subgraph isomorphism imposes too strict a constraint on the topology of the graphs matched.

When graph simulation or its extensions [19, 18] are adopted, *all* four biologists in $G_1$ are matches for Bio in $Q_1$. However, $Bio_1$ and $Bio_2$ are recommended by either HR only or by SE only in $G_1$, and $Bio_3$ is recommended by neither an HR nor an SE. Hence these are not the ones that the headhunter really wants. Only Bio$_4$ satisfies all these conditions and makes a good candidate.

This tells us that simulation and its extensions [19, 18] do not preserve the structural properties in graph pattern matching and therefore, may return excessive "matches" that one does not want. Indeed, observe the following.

*Topological structure*. (a) While $Q_1$ is a connected graph, $G_1$ is disconnected, but $G_1$ matches $Q_1$ via simulation. (b) Node Bio in $Q_1$ has three "parents", but it matches nodes Bio$_1$ and Bio$_2$ in $G_1$ that have a single "parent" each. (c) The directed cycle with only two nodes AI and DM in $Q_1$ matches the long cycle consisting of AI$_1$, DM$_1$, ..., AI$_k$, DM$_k$, AI$_1$ in $G_1$, and the undirected cycle with nodes HR, SE and Bio in $Q_1$ matches the tree rooted at HR$_1$ in $G_1$.

*Match results*. The match relation of simulation, when represented as a result graph as suggested in [19], is the entire graph $G_1$. In general, the result graphs are often large when matching is performed on real-life networks, *e.g.,* LinkedIn [1], which has 19.5M users and yields a graph of 100GB in size. These make it hard to analyze the match results. □

These suggest that we revise the notion of graph simulation to strike a balance between its computational complexity and its ability to capture the topology of graphs. Indeed, graph simulation was proposed for process algebra to mimic steps of a process [29]. To make practical use of it in graph pattern matching, we need to enhance it by incorporating more topological structure of graphs.

**Contributions & Roadmap.** We introduce a revision of graph simulation that preserves the topology of graphs and has the same complexity as extensions [19, 18] of simulation.

(1) We propose a revision of graph simulation [29] (Section 2), referred to as *strong simulation*, by enforcing two conditions: (a) the duality to preserve the parent relationships and (b) the locality to eliminate excessive matches. For example, matching $Q_1$ on $G_1$ of Fig. 1 via strong simulation finds Bio$_4$ as the only match for Bio in $Q_1$.

(2) We identify a set of criteria for topology preservation, and show that strong simulation preserves the topology of pattern graphs and data graphs (Section 3). We also prove that the number of matches via strong simulation is linear in the size of the data graph rather than exponential for subgraph isomorphism, and each match has a diameter bounded by the diameter of the pattern graph. Hence strong simulation indeed rectifies the problems of subgraph isomorphism and simulation. Moreover, we show that slight extensions to the notion make graph pattern matching intractable.

(3) We show that strong simulation retains the same complexity as earlier extensions [19, 18] by providing a *cubic*-time computation algorithm (Section 4). We also develop effective optimization techniques, notably a quadratic-time algorithm to minimize strong simulation queries.

In addition, we show that the locality of strong simulation allows us to develop a simple yet effective algorithm to find matches in distributed graphs. To the best of our knowledge, this is among the first distributed algorithms for graph pattern matching, for which the need is evident when processing massive graphs (see *e.g.,* [15, 21, 28]).

(4) Using both real-life data (Amazon and YouTube) and synthetic data, we conduct an extensive experimental study (Section 5). We find that our algorithms for strong simulation scale well with large data graphs (*e.g.,* with $1.5 \times 10^8$ nodes). They are able to identify sensible matches that subgraph isomorphism fails to catch, and to eliminate excessive matches of graph simulation that do not make sense. We find 70%-80% matches found by strong simulation are those found by subgraph isomorphism, while only 25%-38% for graph simulation. We also find that our optimization techniques are effective, reducing 1/3 of running time in average.

We contend that strong simulation provides a promising model for graph pattern matching in emerging applications. Indeed, (1) in contrast to subgraph isomorphism, strong simulation is in cubic-time rather than NP-complete, and moreover, due to its locality, it yields a set of matches with cardinality linear in the size of the data graph rather than exponential, where each match is bounded by the diameter of the pattern graph. (2) As opposed to graph simulation, it captures the topology of patterns in its matches, such as parents, connectivity and cycles, by enforcing the duality and locality on matches, while it retains the same complexity as simulation. (3) Unlike simulation, the locality of strong simulation makes it possible to efficiently conduct pattern matching on distributed graphs. (4) As will be seen in Section 3, minor extensions to strong simulation would make graph pattern matching an intractable problem. In other words, strong simulation strikes a balance between the complexity and the capability to capture graph topology.

**Related work**. There has been a host of work on pattern matching via subgraph isomorphism (*e.g.,* [32, 33, 35]; see [4, 20] for surveys). In light of its intractability, approximate matching has been studied to find inexact solutions, which allows node/edge mismatches [4, 32]. This work differs from approximate matching in that no node/edge mismatches are allowed, and that the number of matches via strong simulation is linear in the size of the data graph rather than exponential for (approximate) subgraph isomorphism.

Closer to this work are bounded simulation [19] and graph pattern queries of [18]. The former extends simulation [29] by allowing bounds on the number of hops in pattern graphs, and the latter further extends [19] by incorporating regular expressions as edge constraints on pattern graphs. Pattern matching via these extensions are in cubic-time [18, 19]. As remarked earlier, these notions of simulation may fail to capture the topology of graphs, and yield false matches or too large a match relation. These are precisely the problems that strong simulation aims to rectify, by imposing additional constraints (duality and locality) on graph simulation.

Schema extraction is to discover the implicit structure of semi-structured data, which has no schema predefined. It has proved effective in query formulation and optimization [2, 22]. Schema of semi-structured data is often extracted via a mild generalization of simulation that deals with labeled edges [2]. Nevertheless, topology preservation is not an issue in schema extraction, and no previous work there has studied how simulation should be refined to capture topology.

Query minimization, as a classical optimization technique, has been well studied for SQL [3], XPath (*e.g.,* [10]), graph simulation [8] and graph pattern queries [18]. This work explores it for pattern matching via strong simulation.

Distributed query processing has been studied for relational data [26] and XML [9, 11]. There has also been recent work on distributed graph processing to manage large-scale graphs [15, 21, 28]. However, to the best of our knowledge, no previous work has studied distributed computation of graph simulation [29] and its extensions [19, 18], not to mention strong simulation proposed in this work.

## 2. Strong Simulation

In this section, we first present basic notations of graphs. We then introduce the notion of strong simulation.

### 2.1 Preliminaries

We specify both data graphs and pattern graphs as follows. Let $\Sigma$ be a (possibly infinite) set of labels.

*Graphs*. A *node-labeled directed graph* (or simply a *graph*) is defined as $G(V, E, l)$, where (1) $V$ is a finite set of nodes; (2) $E \subseteq V \times V$ is a finite set of edges, in which $(u, u')$ denotes an edge from nodes $u$ to $u'$; and (3) $l$ is a labeling function that maps each node $u$ in $V$ to a label $l(u)$ in $\Sigma$. We denote $G$ as $(V, E)$ when it is clear from the context.

Intuitively, the function $l()$ specifies node attributes, *e.g.*, keywords, blogs, comments, ratings, names, emails, companies [5]; and the label set $\Sigma$ denotes all such attributes.

We next review some basic notations of graphs.

*Subgraphs*. Graph $H(V_s, E_s, l_H)$ is a *subgraph* of graph $G(V, E, l_G)$, denoted as $G[V_s, E_s]$, if (1) for each node $u \in V_s$, $u \in V$ and $l_H(u) = l_G(u)$, and (2) for each edge $e \in E_s$, $e \in E$. That is, subgraph $G[V_s, E_s]$ only contains a subset of nodes and a subset of edges of $G$.

*Paths*. A *directed* (resp. *undirected*) *path* $\rho$ is a sequence of nodes $v_1, \ldots, v_n$ such that $(v_i, v_{i+1})$ (resp. either $(v_i, v_{i+1})$ or $(v_{i+1}, v_i)$) is an edge in $G$ for $i \in [1, n-1]$. The *length* of $\rho$ is the number of edges in $\rho$. Abusing notations for trees, we refer to $v_{i+1}$ as a *child* of $v_i$ (or $v_i$ as a *parent* of $v_{i+1}$).

A *directed* (resp. *undirected*) *cycle* in a graph is a *directed* (resp. *undirected*) path with $v_1 = v_n$, having no repeated nodes other than the start and end nodes $v_1$ and $v_n$.

A graph is *connected* if for each pair of nodes, there exists an undirected path in the graph.

*Connected components*. A *connected component* of a graph is a subgraph in which any two nodes are connected to each other by undirected paths, and which is connected to no additional nodes. A graph that is itself connected has exactly one connected component, consisting of the entire graph.

*Distance and diameter*. Given two nodes $u, v$ in a graph $G$, the *distance* from $u$ to $v$, denoted by $\mathsf{dist}(u, v)$, is the length of the shortest *undirected path* from $u$ to $v$ in $G$.

The *diameter* of a connected graph $G$, denoted by $d_G$, is the longest shortest distance of all pairs of nodes in $G$, *i.e.*, $d_G = \mathsf{max}(\mathsf{dis}(\mathsf{u}, \mathsf{v}))$ for all nodes $u, v$ in $G$.

We assume *w.l.o.g.* that pattern graphs are connected.

### 2.2 The Definition of Strong Simulation

We define strong simulation by enforcing two conditions on simulation [29]: duality and locality. As will be seen in Sections 3 and 4, these conditions capture the topology of graphs and eliminate excessive matches to a maximum extent, while retaining a low PTIME computational complexity.

Consider a pattern graph $Q(V_q, E_q)$ and a data graph $G(V, E)$. To define strong simulation, we first review the notion of graph simulation [29]. Graph $G$ matches pattern $Q$ via *graph simulation*, denoted by $Q \prec G$, if there exists a binary *match relation* $S \subseteq V_q \times V$ such that (1) for each $(u, v) \in S$, $u$ and $v$ have the same label, *i.e.*, $l_Q(u) = l_G(v)$; and (2) for each node $u$ in $V_q$, there exists $v$ in $V$ such that (a) $(u, v) \in S$, and (b) for each edge $(u, u') \in E_q$, there exists an edge $(v, v')$ in $E$ such that $(u', v') \in S$.

Intuitively, simulation preserves the labels and the child relationship of a graph pattern in its match. Simulation was proposed for the analyses of programs [29], and studied for schema extraction from semi-structured data [2]. Simulation and its extensions were recently introduced for social networks [6], and for graph pattern matching [19, 18] due to its low PTIME computational complexity [24].

To capture graph topology, we extend simulation by enforcing duality, to preserve the parent relationship as well.

**Dual simulation**. Pattern graph $Q$ *matches* data graph $G$ via *dual simulation*, denoted by $Q \prec_D G$, if $Q \prec G$ with a binary *match relation* $S \subseteq V_q \times V$, and moreover, for each pair $(u, v) \in S$ and each edge $(u_2, u)$ in $E_q$, there exists an edge $(v_2, v)$ in $E$ with $(u_2, v_2) \in S$.

Intuitively, dual simulation enhances graph simulation by imposing an additional condition, to preserve both child and parent relationships (downward and upward mappings).

While there may be multiple matches in a graph $G$ for a pattern $Q$, there exists a unique *maximum* match $S_M$ in $G$ for $Q$ such that for any match $S$ in $G$ for $P$, $S \subseteq S_M$.

**Lemma 1:** *For any pattern graph $Q$ and data graph $G$ with $Q \prec_D G$, there is a unique maximum match relation.* □

**Locality**. To define the locality, we need some notions.

*Balls*. For a node $v$ in a graph $G$ and a non-negative integer $r$, the *ball* with *center* $v$ and *radius* $r$ is a subgraph of $G$, denoted by $\hat{G}[v, r]$, such that (1) for all nodes $v'$ in $\hat{G}[v, r]$, the shortest distance $\mathsf{dist}(v, v') \leq r$, and (2) it has exactly the edges that appear in $G$ over the same node set.

We define the locality by requiring matches to be within a ball of a certain radius. Indeed, as observed in [7], when social distance increases, the closeness of relationships decreases and the relationships may become irrelevant. Hence it often suffices in practice to consider only those matches of a pattern that fall in a small ball. To formalize this, we use the notion of match graphs of a pattern, given as follows.

*Match graphs*. Consider a relation $S \subseteq V_q \times V$. The *match graph w.r.t. S* is a subgraph $G[V_s, E_s]$ of $G$, in which (1) a node $v \in V_s$ iff it is in $S$, and (2) an edge $(v, v') \in E_s$ iff there exists an edge $(u, u')$ in $Q$ with $(u, v) \in S$ and $(u', v') \in S$.

We are now ready to define strong simulation.

**Strong simulation**. Pattern graph $Q$ *matches* data graph $G$ via *strong simulation*, denoted by $Q \prec^L_D G$, if there exist a node $v$ in $G$ and a connected subgraph $G_s$ of $G$ such that (1) $Q \prec_D G_s$, with the maximum match relation $S$; (2) $G_s$ is exactly the match graph *w.r.t. S*, and (3) $G_s$ is contained in the ball $\hat{G}[v, d_Q]$, where $d_Q$ is the diameter of $Q$.

We refer to $G_s$ as a *perfect* subgraph of $G$ *w.r.t. Q*.

Intuitively, a match $G_s$ of pattern $Q$ is required to satisfy the following conditions: (1) it preserves both the child and parent relationships of $Q$ (condition 1 above); (2) the nodes and edges needed to match $Q$ are all contained in a ball of a radius decided by the diameter of $Q$ (conditions 2 and 3); this rules out excessively large matches. As will be seen shortly, these conditions are justified for preserving graph topology and retaining low computational complexity.

**Example 2:** Consider pattern graph $Q_1$ and data graph $G_1$ of Fig. 1. Observe the following. (1) No subgraph of $G_1$ is isomorphic to $Q_1$. Indeed, there exists no directed cycle in $G_1$ that matches the direct cycle $\mathsf{DM}, \mathsf{AI}, \mathsf{DM}$ in $Q_1$.

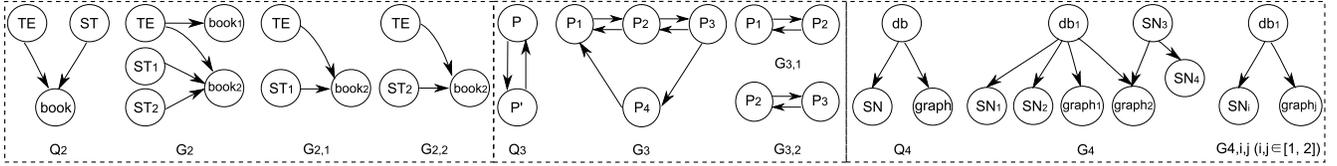

Figure 2: Strong simulation

(2) When simulation is adopted, the entire data graph $G_1$ is included in the match relation, which maps HR, SE, Bio, DM and AI in $Q_1$ to $\{HR_1, HR_2\}$, $\{SE_1, SE_2\}$, $\{Bio_1, Bio_2, Bio_3, Bio_4\}$, $\{DM'_1, DM'_2, DM_1, \ldots, DM_k\}$ and $\{AI'_1, AI'_2, AI_1, \ldots, AI_k\}$ in $G_1$, respectively.

(3) When it comes to strong simulation, the connected component $G_c$ of $G_1$ that contains $Bio_4$ is the only match, which maps HR, SE, Bio, DM and AI in $Q_1$ to $\{HR_2\}$, $\{SE_2\}$, $\{Bio_4\}$, $\{DM'_1, DM'_2\}$ and $\{AI'_1, AI'_2\}$ in $G_1$, respectively. Indeed, one can verify the following: (1) $Q_1 \prec_D G_c$, and in its match relation, Bio in $Q_1$ can only be mapped to $Bio_4$ in $G_1$; and (b) the ball with center $Bio_4$ and radius 3 (the diameter of $Q_1$) is exactly $G_c$. As opposed to simulation, the cycle $AI_1, DM_1, \ldots, AI_k, DM_k, AI_1$ in $G_1$ is not part of the match. Indeed, this cycle is irrelevant and thus should be left out.

As another example, consider pattern graphs $Q_2$, $Q_3$, $Q_4$ and data graphs $G_2$, $G_3$, $G_4$ shown in Fig. 2.

(4) Pattern $Q_2$ is to find a book recommended by both students (ST) and teachers (TE). When simulation is used, both $book_1$ and $book_2$ in $G_2$ are returned as matches, while $book_1$ is obviously not a good option. When strong simulation is adopted, $book_2$ is the only match by the *duality*, in a single match graph (union of $G_{2,1}, G_{2,2}$ in Fig. 2). When it comes to subgraph isomorphism, it returns two match graphs ($G_{2,1}, G_{2,2}$) instead of one, with $book_2$ as the match.

(5) Pattern $Q_3$ is to find people (P and P') who recommend each other. When simulation or dual simulation is used, all people ($P_1$, $P_2$, $P_3$ and $P_4$) in $G_3$ are found as matches, while $P_4$ is obviously not a good choice. When strong simulation is adopted, $P_1$, $P_2$ and $P_3$ are the only matches by the *locality*, in a single match graph (union of $G_{3,1}, G_{3,2}$ in Fig. 2). These are also the matches found via subgraph isomorphism, in two match graphs ($G_{3,1}, G_{3,2}$) instead of a single one.

(6) Pattern $Q_4$ is looking for papers on social networks (SN) cited by papers on databases (db), which in turn cite papers on graph theory (graph). When simulation is used, all papers on SN ($SN_1$, $SN_2$, $SN_3$ and $SN_4$) in $G_4$ are matches, while $SN_3$ and $SN_4$ are obviously excessive matches. When strong simulation is adopted, $SN_1$ and $SN_2$ are the only matches due to the *duality*, returned in a single match graph (union of $G_{4,i,j}$ with $i, j \in [1, 2]$ in Fig. 2). These are also the matches found by subgraph isomorphism, yet returned in four match graphs ($G_{4,i,j}$ for $i, j \in [1, 2]$) instead of one. □

**Semantics**. Strong simulation is to find, given any pattern graph $Q$ and data graph $G$, the set of the *maximum* perfect subgraph $G_s$ in each ball such that $Q \prec_D G_s$.

By Lemma 1, one can verify the following, which assures that dual simulation and strong simulation are well defined.

**Theorem 1:** *For any pattern graph $Q$ and data graph $G$ such that $Q \prec_D^L G$, there exists a unique set of maximum perfect subgraphs for $Q$ and $G$.* □

| Notations | Semantics |
|---|---|
| $G[V_s, E_s]$ | Subgraph of $G$ with node and edge sets $V_s$, $E_s$ |
| $\hat{G}[v, r]$ | A ball in a graph $G$ with center $v$ and radius $r$ |
| $\triangleleft, \prec$ | Subgraph isomorphism and graph simulation |
| $\prec_D, \prec_D^L$ | Dual simulation and strong simulation |

Table 1: Summary of notations

**Remark**. (1) Duality and locality are also imposed by subgraph isomorphism, but not by simulation. (2) One can readily extend strong simulation by supporting bounds on the number of hops and regular expressions as edge constraints on pattern graphs, along the same lines as [19, 18]. We defer this to the full report to simplify the discussion.

We summarize notations in Table 1, in which we use $Q \triangleleft G$ to denote that $Q$ matches $G$ via subgraph isomorphism.

## 3. Properties of Strong Simulation

Below we first identify a set of criteria for topology preservation in pattern matching and for bounded match results. Based on the criteria we then evaluate strong simulation, dual simulation, subgraph isomorphism and graph simulation. Finally, we explore possible extensions to strong simulation and show that they lead to intractable problems.

Consider a connected pattern graph $Q = (V_q, E_q)$ with diameter $d_Q$ and a data graph $G = (V, E)$.

### 3.1 Fundamental Properties

First, one can readily verify that subgraph isomorphism is a stronger notion than the other three, followed by strong simulation, dual simulation and graph simulation in this order. Intuitively, subgraph isomorphism preserves most topological structures between data graphs and pattern graphs.

**Proposition 1:** *(1) If $Q \triangleleft G$, then $Q \prec_D^L G$; (2) if $Q \prec_D^L G$, then $Q \prec_D G$; and (3) if $Q \prec_D G$, then $Q \prec G$.* □

We next take a closer look at what structures are preserved by these matching notions, by giving a set of criteria.

*(1) Children*. If a node $u$ in the pattern graph $Q$ matches node $v$ in the data graph $G$, then each child of $u$ must match a child of $v$. All these notions preserve the child relationship.

*(2) Parents*. If a node $u$ in $Q$ matches node $v$ in $G$, then each parent of $u$ also matches a parent of $v$. One can verify that subgraph isomorphism, strong simulation and dual simulation preserve the parent relationship, but simulation does not. A counterexample for simulation is given in Fig. 1.

*(3) Connectivity*. A connected pattern graph only matches a connected subgraph in the data graph.

As shown in Example 1, connected $Q_1$ may match a disconnected data graph $G_1$ via graph simulation. Dual simulation prevents this, as shown below. From this it follows that the stronger notions subgraph isomorphism and strong simulation also preserve the connectivity.

**Theorem 2:** *If $Q \prec_D G$, then for any connected component*

| matching | children | parents | connectivity | cycles | locality | matches | bisimilar&b'ed-cycle |
|---|---|---|---|---|---|---|---|
| $\prec$ | ✓ | × | × | ✓(directed), ×(undirected) | × | ✓ | × |
| $\prec_D$ | ✓ | ✓ | ✓ | ✓(directed & undirected) | × | × | × |
| $\prec_D^L$ | ✓ | ✓ | ✓ | ✓(directed & undirected) | ✓ | ✓ | × |
| $\triangleleft$ | ✓ | ✓ | ✓ | ✓(directed & undirected) | ✓ | × | ✓ |

Table 2: Topology preservation and bounded matches

$G_c$ of the match graph w.r.t. the maximum match relation of $Q$ and $G$, (1) $Q \prec_D G_c$, and (2) $G_c$ is exactly the match graph w.r.t. the maximum match relation of $Q$ and $G_c$. □

Since $G_c$ is a connected component of the match graph and $Q$ is assumed connected, all "relevant nodes" are in $G_c$.

*(4) Cycles.* An undirected (resp. directed) cycle in $Q$ must match an undirected (resp. directed) cycle in $G$.

We show that graph simulation preserves directed cycles, and hence so do the other three matching notions.

**Proposition 2:** *If $Q \prec G$ and there is a directed cycle in $Q$, then there must exist a matched directed cycle in the match graph w.r.t. any match relation of $Q$ and $G$.* □

However, as shown in Example 1, graph simulation may match an undirected cycle in a pattern with a tree in a data graph. In contrast, dual simulation (and subgraph isomorphism and strong simulation) preserve undirected cycles.

**Theorem 3:** *If $Q \prec_D G$ and there is an undirected cycle in $Q$, then there must exist a matched undirected cycle in the match graph w.r.t. any match relation of $Q$ and $G$.* □

*(5) Locality.* The diameter of a matched subgraph in $G$ must be bounded by a function in the size of the pattern graph. This allows us to check a match locally, by inspecting only its neighborhood of a bounded diameter.

Strong simulation has the locality property, and so does subgraph isomorphism. In contrast, neither simulation nor dual simulation has the locality (see Examples 1 and 2).

**Proposition 3:** *If $Q \prec_D^L G$, then for all perfect subgraphs $G_s$ of $G$, the diameter of $G_s$ is bounded by $2 * d_Q$, where $d_Q$ is the diameter of $Q$.* □

*(6) Bounded matches.* There should be a bounded number of matches, and each match is small enough to inspect. As remarked earlier, subgraph isomorphism may yield exponentially many matched subgraphs. While simulation and dual simulation return a single match relation, the relation is often too large to understand. In contrast, strong simulation strikes a balance: the number of matches is bounded by the number of nodes in the data graph, and each matched subgraph has a bounded diameter (Proposition 3).

**Proposition 4:** *The number of maximum perfect subgraphs of $G$ is bounded by the number of nodes in $G$.* □

These results are summarized in Table 2. They tell us that strong simulation preserves much more topological structures between pattern graphs and data graphs than graph simulation, and moreover, possesses the locality property.

### 3.2 In Search for Tractable Boundary in Matching

One might want to find a notion of graph pattern matching that preserves maximum graph topology, and characterize PTIME along the same lines as how Fagin's theorem characterizes NP [30]. This is, however, very challenging. Indeed, as observed in [23], in graph theory Fagin's theorem implies that "if no logic captures PTIME, then PTIME ≠ NP".

Below we present two negative results: extending strong simulation makes its computation from PTIME to NP-hard.

*Bounded cycles.* Given a pattern graph $Q$ and a data graph $G$ such that $Q \prec G$ with the maximum match relation $S$, the *bounded cycle problem* is to determine whether the longest cycle in the match graph w.r.t. $S$ is bounded by the longest one in $Q$. Obviously bounded cycle is a desirable locality property that one would have wanted to further impose on strong simulation. Unfortunately, this additional condition would make pattern matching intractable.

**Theorem 4:** *The bounded cycle problem is coNP-hard even when pattern graphs contain a single cycle.* □

*Bisimulation.* One might be tempted to use graph bisimulation [29] rather than graph simulation in graph pattern matching. A pattern graph $Q$ *matches* a graph $G_s$ via *bisimulation*, denoted by $Q \sim G_s$, if $Q \prec G_s$ with the maximum match relation $S$ and $G_s \prec Q$ with the inverse $S^-$ of $S$ as its maximum match relation. Pattern matching via bisimulation is to find all subgraphs $G_s$ of a graph $G$ such that $Q \sim G_s$. Clearly bisimulation preserves more topological structures than simulation. Indeed, it is a notion stronger than simulation but weaker than isomorphism.

However, pattern matching via bisimulation becomes intractable. Indeed, subgraph bisimulation is NP-hard [17], although graph bisimulation is solvable in PTIME [29]. In contrast, subgraph simulation is equivalent to graph simulation, *i.e.*, checking whether there exists a subgraph $G_s$ of $G$ such that $Q \prec G_s$ is the same as checking whether $Q \prec G$.

## 4. An Algorithm for Strong Simulation

We next show that graph pattern matching via strong simulation retains the same complexity as earlier extensions [19, 18] of simulation, while it is able to preserve graph topology better. The main results of this section are as follows.

**Theorem 5:** *For any pattern graph $Q$ and data graph $G$, it takes cubic time to check whether $Q \prec_D^L G$, and to find the set of maximum perfect subgraphs of $G$ w.r.t. $Q$.* □

**Theorem 6:** *For any pattern graph $Q$ with diameter $d_Q$, it takes quadratic time to find a minimum pattern graph $Q_m$ such that $Q_m$ and $Q$ find the same result on any data graph by using $d_Q$ as the radius of balls, via strong simulation.* □

We first prove Theorem 5 by providing a cubic-time algorithm for computing strong simulation. We then show Theorem 6 by proposing optimization techniques. Finally, we briefly discuss how the locality of strong simulation allows us to conduct pattern matching on distributed graphs.

### 4.1 A Cubic-time Algorithm

**Algorithm**. The algorithm, refereed to as Match, is shown in Fig. 3. Given a pattern graph $Q$ and a data graph $G$, it returns the set of perfect subgraph $G_s$ by inspecting those balls of radius $d_Q$ centered at each node $w$ of $G$.

To present Match, we first describe its procedures.

*Input:* Pattern graph $Q$ with diameter $d_Q$, data graph $G(V, E)$.
*Output:* The set $\Theta$ of maximum perfect subgraphs of $G$ w.r.t. $Q$.

1. $\Theta := \emptyset$;
2. **for each** ball $\hat{G}[w, d_Q]$ in $G$ **do**
3.    $S_w := \mathsf{DualSim}(Q, \hat{G}[w, d_Q])$;
4.    $G_s := \mathsf{ExtractMaxPG}(Q, \hat{G}[w, d_Q], S_w)$;
5.    **if** $G_s \neq nil$ **then** $\Theta := \Theta \cup \{G_s\}$;
6. **return** $\Theta$.

**Procedure** $\mathsf{DualSim}(Q, \hat{G}[w, d_Q])$

*Input:* Pattern graph $Q(V_q, E_q)$ and ball $\hat{G}[w, d_Q]$.
*Output:* The maximum match relation $S_w$ of $Q$ and $\hat{G}[w, d_Q]$.

1. **for each** $v \in V_q$ **do**
2.    $\mathsf{sim}(v) := \{u \mid u \text{ is in } \hat{G}[w, d_Q] \text{ and } l_Q(u) = l_G(v)\}$;
3. **while** there are changes **do**
4.    **for each** edge $(v, v')$ in $E_Q$ and **each** node $u \in \mathsf{sim}(v)$ **do**
5.      **if** there is no edge $(u, u')$ in $\hat{G}[w, d_Q]$ with $u' \in \mathsf{sim}(v')$
6.      **then** $\mathsf{sim}(v) := \mathsf{sim}(v) \setminus \{u\}$;
7.    **for each** edge $(v', v)$ in $E_Q$ and **each** node $u \in \mathsf{sim}(v)$ **do**
8.      **if** there is no edge $(u', u)$ in $\hat{G}[w, d_Q]$ with $u' \in \mathsf{sim}(v')$
9.      **then** $\mathsf{sim}(v) := \mathsf{sim}(v) \setminus \{u\}$;
10.   **if** $\mathsf{sim}(v) = \emptyset$ **then return** $\emptyset$;
11. $S_w := \{(v, u) \mid v \in V_q, u \in \mathsf{sim}(v)\}$;
12. **return** $S_w$.

**Procedure** $\mathsf{ExtractMaxPG}(Q, \hat{G}[w, d_Q], S_w)$

*Input:* Pattern $Q$, ball $\hat{G}[w, d_Q]$, maximum match relation $S_w$.
*Output:* The maximum perfect subgraph $G_s$ in $\hat{G}[w, d_Q]$ if any.
1. **if** $w$ does not appear in $S_w$ **then return** $nil$;
2. Construct the matching graph $G_m$ w.r.t. $S_w$;
3. **return** the connected component $G_s$ containing $w$ in $G_m$.

**Figure 3: Algorithm** $\mathsf{Match}$

*Procedure* $\mathsf{DualSim}$. It takes as input pattern graph $Q(V_q, E_q)$ and ball $\hat{G}[w, d_Q]$ with center $w$ and radius $d_Q$, and finds the maximum match relation $S_w$ of $Q$ and $\hat{G}[w, d_Q]$. For each node $v$ in $V_q$, it first computes the set $\mathsf{sim}(v)$ of candidate matches $u$ in the ball with the same node label, i.e., $l_{V_q}(u) = l_V(v)$ (lines 1–2). Then the procedure repeatedly removes nodes from $\mathsf{sim}(v)$ for each node $v$ in $Q$ (lines 3–10). A node $u$ is removed from $\mathsf{sim}(v)$ unless (1) if there is a parent node $v'$ of $v$, then there exists a parent node $u \in \mathsf{sim}(v')$; and (2) if there is a child node $v'$ of $v$, then there exists a child node $u \in \mathsf{sim}(v')$. Finally, $S_w$ is returned (lines 11–12).

*Procedure* $\mathsf{ExtractMaxPG}$. It takes as input a pattern graph $Q$, ball $\hat{G}[w, d_Q]$, and the maximum match relation $S_w$. It finds the perfect subgraph $G_s$ in the ball if there exists one. By Theorem 2, the procedure simply finds the connected component containing $w$ in the match graph w.r.t. $S_w$.

*Algorithm* $\mathsf{Match}$. We are now ready to present $\mathsf{Match}$. For each node $w$ in the data graph $G$, (1) it computes the maximum match relation $S_w$ of $Q$ and the ball $\hat{G}[w, d_Q]$ by invoking $\mathsf{DualSim}$ (line 2); (2) it finds the perfect subgraph $G_s$ in $\hat{G}[w, d_Q]$ via $\mathsf{ExtractMaxPG}$ (line 3); and (3) $G_s$ is added to the set $\Theta$ if it exists (line 4). After all balls in $G$ are checked, it returns the set $\Theta$ of perfect subgraphs (line 5).

**Example 3:** Consider pattern graph $Q_1$ ($d_{Q_1} = 3$) and the ball with center $\mathsf{Bio}_4$ and radius $= 3$ in data graph $G_1$ of Fig 1. Note that the ball is exactly the connected component $G_c$ with node $\mathsf{Bio}_4$ in $G_1$. We show how Algorithm $\mathsf{Match}$ works on $Q_1$ and $G_c$. Initially, HR, SE, Bio, AI and DM in $Q_1$ match $\{\mathsf{HR}_2\}$, $\{\mathsf{SE}_2\}$, $\{\mathsf{Bio}_4,\}$, $\{\mathsf{AI}'_1, \mathsf{AI}'_2\}$ and $\{\mathsf{DM}'_1, \mathsf{DM}'_2\}$ in $G_c$, respectively (lines 1-2, $\mathsf{DualSim}$). The algorithm finds

*Input:* Pattern graph $Q = (V_q, E_q, l_Q)$.
*Output:* A minimized equivalent pattern graph $Q_m$ of $Q$.

1. Compute the maximum match relation $S$ of $Q \prec_D Q$;
2. Compute equivalent classes of nodes in $Q$ based on $S$;
3. Create a node for each equivalent class in $Q_m$;
4. Connect different equivalent classes by necessary edges in $Q_m$;
5. **return** $Q_m$.

**Figure 4: Algorithm** $\mathsf{minQ}$

no nodes to be removed from $\mathsf{sim}(u)$ for all nodes $u$ in $Q_1$ in this case (lines 3-10, $\mathsf{DualSim}$). Hence $\mathsf{Match}$ returns $G_c$ as the perfect subgraph in the ball (line 6, $\mathsf{Match}$). □

*Correctness & Complexity.* The correctness of algorithm is assured by the following. (1) There is at most one perfect subgraph in each ball of $G$ (Theorem 1). (2) Procedure $\mathsf{ExtractMaxPG}$ returns the perfect graph in ball $\hat{G}[v, d_Q]$, by Theorem 2. (3) The correctness of $\mathsf{DualSim}$ can be verified along the same lines as its counterpart for simulation [24].

It takes $\mathsf{BuildBall}$ $O(|V| + |E|)$ time to build a ball $\hat{G}[v, d_Q]$ by using the BFS method [16]. For each ball, $\mathsf{ExtractMaxPG}$ finds its perfect subgraph in $O(|V|)$ time since finding pairwise disconnected components is linear-time equivalent to finding strongly connected components, which is in linear time [13]. By leveraging the algorithm developed in [24], $\mathsf{DualSim}$ can be done in $O((|V_q| + |E_q|)(|V| + |E|))$ time. Thus $\mathsf{Match}$ is in $O(|V|(|V| + (|V_q| + |E_q|)(|V| + |E|)))$ time.

### 4.2 Optimization Techniques

We next present optimization techniques for algorithm $\mathsf{Match}$, by means of query minimization, dual simulation filtering and connectivity pruning.

**Query minimization.** We first explore query minimization, which is important for any query language [3].

We say that two pattern graphs $Q$ and $Q'$ are *equivalent*, denoted by $Q \equiv Q'$, iff they return the same result on any data graph. A pattern graph $Q$ is *minimum* if it has the least size $|Q|$ (the number of nodes and the number of edges) among all equivalent pattern graphs.

By the complexity analysis of algorithm $\mathsf{Match}$, smaller pattern graphs lead to better performance.

Theorem 6 follows from Lemmas 2 and 3 given below.

**Lemma 2:** *For any pattern graph, (1) there exists a unique (up to isomorphism) minimum equivalent pattern graph, via dual simulation, that finds the same maximum match relation on any data graph; and (2) there exists a quadratic time algorithm to find its minimum equivalent query.* □

**Lemma 3:** *When fixing the radius of balls in strong simulation, two pattern graphs are equivalent via strong simulation iff they are equivalent via dual simulation.* □

Leveraging these, Algorithm $\mathsf{Match}$ can be improved as follows. Given query graph $Q$, we first compute its minimum equivalent query graph $Q_m$, and then we compute strong simulation w.r.t. $Q_m$ and diameter $d_Q$.

*Algorithm.* As a proof of Lemma 2, we present Algorithm $\mathsf{minQ}$ for minimizing graph patterns, shown in Fig. 4. It takes as input a pattern graph $Q$, and returns a minimum equivalent pattern $Q_m$ of $Q$, via dual simulation. For any pattern graph $Q$, it first computes the maximum match relation $S$ by treating $Q$ as both a pattern graph and a data graph (line 1). It then computes equivalent classes for nodes in $Q$ such that nodes $u$ and $v$ are in the same class iff both

---

*Input:* Pattern $Q$, relation $S$ w.r.t. $Q \prec_D G$, ball $\hat{G}[w, d_Q]$.
*Output:* The maximum perfect subgraph of $\hat{G}[w, d_Q]$ w.r.t. $Q$.
1. $S_w :=$ project $S$ onto $\hat{G}[w, d_Q]$; $filterSet := \emptyset$;
2. **for each** $(u, v) \in S_w$ such that $v$ is a border node **do**
3.    **if** $\text{succ}(v) \cap \text{sim}(u_1) = \emptyset$ or $\text{pred}(v) \cap \text{sim}(u_2) = \emptyset$
4.      such that $(u, u_1) \in E_q$, $(u_2, u) \in E_q$
5.    **then** $filterSet.\text{push}((u, v))$;
6. **while** ($filterSet \neq \emptyset$) **do**
7.    $(u, v) := filterSet.\text{pop}()$; $S_w := S_w \setminus \{(u, v)\}$;
8.    **for each** $(u_2, u) \in E_q$ **do**
9.      **for each** $v_2 \in \text{pred}(v) \cap \text{sim}(u_2)$ **do**
10.        **if** $\text{succ}(v_2) \cap \text{sim}(u) = \emptyset$ **then**
11.          $filterSet.\text{push}((u_2, v_2))$;
12.    **for each** $(u, u_1) \in E_q$ **do**
13.      **for each** $v_1 \in \text{succ}(v) \cap \text{sim}(u_1)$ **do**
14.        **if** $\text{succ}(v) \cap \text{sim}(u_1) = \emptyset$ **then**
15.          $filterSet.\text{push}((u_1, v_1))$;
16. **if** there exists $u$ in $Q$ such that $\text{sim}(u) = \emptyset$ **then** $S_w := \emptyset$;
17. **return** $\text{ExtractMaxPG}(Q, \hat{G}[w, d_Q], S_w)$.

---

**Figure 5: Algorithm dualFilter**

$(u, v) \in S$ and $(v, u) \in S$ (line 2). Finally, it constructs the minimum equivalent query $Q_m$ as follows (lines 3-4). (a) For each equivalent class eq, it creates a node eq for $Q_m$, and (b) there is an edge (eq, eq′) in $Q_m$ iff there exist nodes $u \in$ eq and $u' \in$ eq′ such that there is an edge $(u, u')$ in $Q$.

**Example 4:** Taking as input the pattern graph $Q_5$ given in Fig. 6(a), Algorithm minQ works as follows. (1) It first computes the maximum match relation $S$ of $Q_5$ and $Q_5$, via dual simulation, yielding $S = \{(R, R), (B_i, B_j), (C_i, C_j), (D_i, D_j)\}$ $(i, j \in [1, 2])$. (2) It then computes five equivalent classes: $\text{eq}_R = \{R\}$, $\text{eq}_A = \{A\}$, $\text{eq}_B = \{B_1, B_2\}$, $\text{eq}_C = \{C_1, C_2\}$, and $\text{eq}_D = \{D_1, D_2\}$. (3) Finally, it constructs the minimum pattern graph $Q_{5,m}$ of $Q_5$, shown in Fig. 6(a): (a) For each equivalent class $\text{eq}_x$, where $x \in \{R, A, B, C, D\}$, it creates a node labeled with $x$; and (b) it creates an edge from node $x$ to $y$ in $Q_{5,m}$ iff there exist node $u \in \text{eq}_x$ and $v \in \text{eq}_y$ such that $(u, v)$ is an edge in $Q_5$. □

*Correctness & Complexity.* The correctness of Algorithm minQ is assured by the following. (1) For any data graph $G$, the match graph w.r.t. the maximum match relation $S$ of $Q$ and $G$ is always the same as the the match graph w.r.t. the maximum match relation $S_m$ of $Q_m$ and $G$. Hence, $Q \equiv Q_m$. (2) $|Q_m| \leq |Q'|$ for any $Q'$ such that $Q' \equiv Q$. (3) For any two minimum equivalent pattern graphs $Q_m$ and $Q'_m$, there is a bijective mapping from $Q_m$ to $Q'_m$ such that (a) for any node $u$ in $Q_m$, $f(u)$ is a node in $Q'_m$ with the same label, and (b) $(u, v)$ is an edge in $Q_m$ iff $(f(u), f(v))$ is an edge in $Q'_m$, e.g., $Q_m$ and $Q'_m$ are equivalent up to isomorphism.

Algorithm minQ is in $O((|V_q|+|E_q|)^2)$ time. Indeed, steps (1), (2) and (3) of minQ take $O((|V_q|+|E_q|)^2)$ time, $O(|V_q|^2)$ time and $O(|E_q|)$ time, respectively.

**Dual simulation filtering.** Our second optimization technique aims to avoid redundant checking of balls in the data graph. Most algorithms of graph simulation (*e.g.,* [24]) recursively refine the match relation by identifying and removing false matches. As observed in [19], it is much easier to deal with node or edge deletions than their insertions. In light of this, we compute the match relation of dual simulation first, and then project the match relation on each ball to compute strong simulation. This both reduces the initial match set $\text{sim}(v)$ for each node $v$ in $Q$ (line 2, Dualsim), and reduces the number of balls (line 2, Match). Indeed, if a node $v$ in $G$ does not match any node in $Q$, then there is no need to consider the ball centered at $v$.

Consider *border* nodes in a ball $\hat{G}[v, r]$, *i.e.,* nodes $u$ with $\text{dist}(v, u) = r$. We refer to those nodes reachable from a border node as *affected* nodes. One can verify the following:

**Proposition 5:** *The removal process on a ball only needs to deal with its border nodes and their affected nodes.* □

This suggests an order to process nodes in $\hat{G}[v, r]$: starting from its border nodes, we inspect affected nodes only following an increasing order based on their distances from border nodes. This minimizes unnecessary computation. Note that the border nodes can be marked when constructing balls. Hence this incurs little extra complexity.

*Algorithm.* To do this, we first compute the match relation $\overline{S_G}$, via dual simulation, over the entire data graph by invoking Procedure DualSim in Fig. 3. We then project $S_G$ onto each ball. When computing the perfect subgraph on a ball, we start with the *border* nodes, and identify invalid matches using Algorithm dualFilter shown in Fig. 5.

We next present Algorithm dualFilter. It takes as input pattern graph $Q$, the maximum match relation $S$ of $Q$ and $G$ that is found via dual simulation, and ball $\hat{G}[w, d_Q]$. It returns the maximum perfect subgraph of $\hat{G}[w, d_Q]$ w.r.t. $Q$. More specifically, dualFilter first projects match relation $S_G$ onto ball $\hat{G}[w, d_Q]$, yielding relation $S_w$ (line 1). It then iteratively marks and removes those invalid matches stored in a queue *filterSet* (lines 2-15), initially empty. To do this, it first inspects those matches in $S_w$ that contain a border node, to find invalid matches (lines 2-4). The invalid matches found are stored in *filterSet* (line 5). It then processes those marked invalid matches one by one (lines 6-15). Each invalid match $(u, v)$ with affected node $v$ is removed from $S_w$ (line 7). The relation $S_w$ is then processed along the same lines as Algorithm Match (lines 8-15), but following the order of invalid matches in *filterSet*. Finally, the algorithm extracts the perfect subgraph by invoking Procedure ExtractMaxPG, and returns the subgraph (line 17).

**Example 5:** We next illustrate how the filtering technique improves the performance of Algorithm Match by considering pattern graph $Q_6$ and data graph $G_6$ given in Fig. 6(b). The maximum match relation $S_{G_6}$ of $Q_6$ and $G_6$ via dual simulation is the set of matches (node pairs) $\{(A, A_2), (A, A_3), (B, B_2), (B, B_3), (C, C)\}$. Hence initially, $\text{sim}(A) = \{A_2, A_3\}$, $\text{sim}(B) = \{B_2, B_3\}$ and $\text{sim}(C) = \{C\}$.

The filtering method then projects the match relation $S_{G_6}$ on each ball, and checks the results. It finds the following.
(i) There exist invalid matches in two balls: $\hat{G}_6(A_1, 3)$ and $\hat{G}_6(B_1, 3)$, by inspecting their border nodes. For $\hat{G}_6(A_1, 3)$, after projecting $S_{G_6}$ on $\hat{G}_6(A_1, 3)$, we get $S_w = \{\text{sim}(A) = \text{sim}(A') = \{A_2\}, \text{sim}(B) = \text{sim}(B') = \{B_2\}\}$. Here $B_2$ is a border node of $\hat{G}[A_1, 3]$. Starting with $B_2$, dualFilter finds that there exist invalid matches; similarly for $\hat{G}_6(B_1, 3)$.
(ii) In contrast, there exist no invalid matches in balls $\hat{G}_6(A_2, 3)$, $\hat{G}_6(B_2, 3)$, $\hat{G}_6(A_3, 3)$, $\hat{G}_6(B_3, 3)$ and $\hat{G}_6(C, 3)$. This is found by inspecting border nodes in each ball. Hence the final match relation in any of these balls is exactly the same as the initial projection of $S_{G_6}$ on the ball.

As a result, only two balls ($\hat{G}_6(A_1, 3)$ and $\hat{G}_6(B_1, 3)$) are really processed by dualFiler, while no more processing is needed for the other five balls. That is, the filtering method prunes unnecessary processing and speeds up Match. □

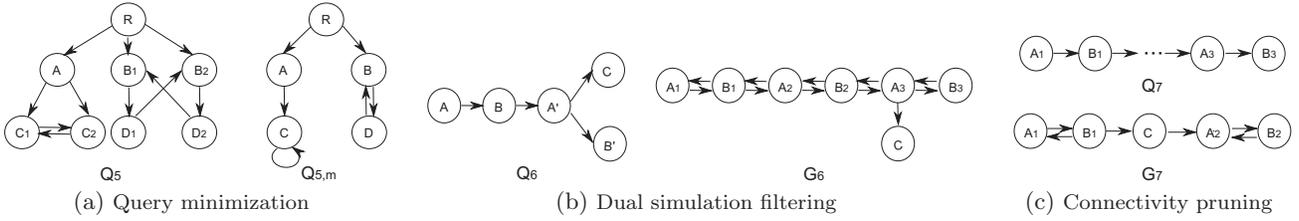
(a) Query minimization  (b) Dual simulation filtering  (c) Connectivity pruning

**Figure 6: Examples for optimization techniques**

*Correctness & Complexity.* The correctness of Algorithm dualFiler is asserted by Proposition 5. For its complexity, observe that it takes $O(|V|(|V| + |E|))$ time to construct all balls, and $O((|V_q| + |E_q|)(|V| + |E|))$ time to compute the maximum match relation of $Q$ and $G$ via dual simulation, along the same lines as Algorithm Match. For each ball $\hat{G}[w, d_Q]$, it takes at most $O((|V_q| + |E_q|)(|V_{\hat{G}[w,d_Q]}| + |E_{\hat{G}[w,d_Q]}|))$ time. Putting these together, dualFilter takes $O(|V|(|V|+(|V_q|+|E_q|)(|V|+|E|)))$ time in total. Although the worst case complexity is the same as the complexity of Match (shown in Fig. 3), as demonstrated by the example and as will be shown by our experimental study, the optimization technique is indeed effective in practice.

**Connectivity pruning.** Theorem 2 tells us that in a ball $\hat{G}[v, r]$, only the connected component containing the ball center $v$ needs to be considered. Hence, those nodes not reachable from $v$ can be pruned early. Our last main optimization technique does precisely this. It reduces the search space for checking dual-simulation, and can be easily incorporated into Algorithm Match, as illustrated below.

**Example 6:** Consider pattern graph $Q_7$ and data graph $G_7$ shown in Fig. 6(c), in which diameters $d_{Q_7} = 5$ and $d_{G_7} = 4$. As $d_{Q_7} > d_{G_7}$, a ball with any center node of $G$ is exactly $G$ itself. When conducting dual simulation of $Q_7$ on ball $\hat{G}_7[A_1, 5]$, for instance, the pruning method first finds an initial sim(v) set for each node $v$ in $Q_7$, by mapping $A_i$ in $Q_7$ to $A_j$ in $\hat{G}_7[A_1, 5]$ ($i \in [1, 3], j \in [1, 2]$). This yields two connected component in $\hat{G}_7[A_1, 5]$: $SC_1$ containing nodes $\{A_1, B_1\}$ and $SC_2$ containing $\{A_2, B_2\}$, in which only $SC_1$ contains the center node $A_1$ (recall the notion of connected graphs from Section 2). By Theorem 2, the pruning method safely removes all those nodes that are not in $SC_1$ from sim(u), for any node $u \in Q_7$, without affecting the final matches. That is, it prunes invalid matches early and hence, improves the performance of algorithm Match. □

We have implemented a version of Match that supports all optimization strategies, referred to as Match$^+$. As will be seen in Section 5, Match$^+$ significantly outperforms Match.

### 4.3 Strong Simulation on Distributed Graphs

When evaluating a query on a large dataset, one wants to partition the data and distribute its fragments to multiple machines, such that the query can be evaluated in parallel, as advocated by, *e.g.,* MapReduce [15]. Moreover, it is common to find real-life datasets already partitioned and distributed. For instance, to find the complete information of a person, one may have to query several social networks (*e.g.,* Facebook, Picassa and Youtube) to collect her data. These highlight the need for developing distributed algorithms for evaluating graph queries. However, as observed in [28], graph algorithms often exhibit poor data locality and hence, may incur prohibitive overhead on network traffic.

We next show that strong simulation demonstrates data locality and hence, allows efficient distributed evaluation.

**Data locality.** Consider a graph $G$ that is partitioned into $(G_1, \ldots, G_k)$ such that each $G_i$ is stored in site $M_i$ for $i \in [1, k]$. We want to evaluate a pattern query $Q$ on $G$, while minimizing unnecessary data shipment from one site to another. This is, however, rather challenging when pattern matching is defined in terms of graph simulation.

**Example 7:** Consider again query graph $Q_1$ and data graph $G_1$ of Fig. 1. Let $G_s$ be the subgraph of $G_1$ by removing the connected component with Bio$_4$ from $G_1$. Suppose that $G_s$ is fragmented and distributed. Then to decide whether $Q_1 \prec G_s$, we have to ship all subgraphs of $G_s$ to a single site to re-assemble $G_s$. Indeed, (1) the match graph of $Q_1$ and $G_s$ via graph simulation is the entire $G_s$; and (2) removing any node or edge from $G_s$ makes $Q_1 \not\prec G_s$. This tells us that it is hard to conduct graph simulation in the distributed setting without incurring high network traffic. □

In contrast, we show that strong simulation has the data locality. Indeed, strong simulation can be computed in the distributed setting, guaranteeing that the total data shipment is bounded by the set of balls $\hat{G}[v, d_Q]$ in $G$ such that $v$ is in some $G_i$ but it has a direct neighbor node not in $G_i$.

**Algorithm.** To verify the data locality of strong simulation, we outline a distributed algorithm for strong simulation.

When a site, referred to as *the coordinator*, receives a pattern graph $Q$, it sends the same $Q$ to each site $M_i$ for $i \in [1, k]$. When a site $M_i$ receives $Q$, it finds those balls $\hat{G}[v, d_Q]$, where $v$ is in $G_i$ but $v$ has a neighbor node in another fragment $G_j$. For such nodes $v$, $M_i$ sends $\hat{G}[v, d_Q]$ to site $M_j$ only if $j < i$. It then invokes algorithm Match to compute the matches of $Q$ in $G_i$, as a partial result $\Theta_i$ of $Q$ in $G$, at site $M_i$. It sends $\Theta_i$ back to the coordinator.

The coordinator monitors messages sent back from all sites $M_i$ for $i \in [1, k]$. When partial results are returned from all the sites, the coordinator assembles their partial results via union, and returns the final result.

One can readily verify that the algorithm is correct, with the bound on network traffic mentioned above. Furthermore, it is generic: it is applicable to any $G$ regardless of how $G$ is partitioned and distributed.

## 5. Experimental Study

We next present an experimental study of strong simulation. Using both real-life social networks and synthetic data, we conducted two sets of experiments to evaluate: (1) the effectiveness of strong simulation vs. conventional subgraph isomorphism [34] and graph simulation [24], and (2) the efficiency of our centralized algorithm Match.

**Experimental setting.** We used the following datasets.

*Real-life graph data.* We used two real-life network datasets.

(1) *Amazon* records a product co-purchasing network with 548,552 product nodes and 1,788,725 product-product directed edges[1]. An edge from products $x$ to $y$ indicates that people buy $y$ with high probability when they buy $x$.

(2) *YouTube* provides a video network with 155,513 video nodes and 3,110,120 video-video directed edges[2]. An edge from videos $x$ to $y$ indicates that if one watches $x$, then he is also very likely to watch $y$.

<u>Synthetic graph generator</u>. We adopted the graph-tool library[3] to produce both pattern and data graphs. It is controlled by three parameters: the number $n$ of nodes, the number $n^\alpha$ of edges, and the number $l$ of node labels. Given $n, \alpha$, and $l$, the generator produces a graph with $n$ nodes, $n^\alpha$ edges, and the nodes are labeled from a set of $l$ labels.

<u>Algorithms</u>. We implemented the following algorithms, all in Python: (1) our algorithms Match and Match$^+$, (2) the graph simulation algorithm of [24], denoted by Sim, (3) the approximate matching algorithm TALE of [32], and (4) an approximate matching algorithm, denoted by MCS, that utilizes the approximation algorithm of [25] for computing maximum common subgraphs. We used the VF2 algorithm [12] for subgraph isomorphism in the igraph package [14].

Consider pattern graph $Q(V_q, E_q)$ and data graph $G(V, E)$. For approximate matching algorithms TALE and MCS, there are essentially $2^{|V|}$ number of subgraphs of $G$ to compare with $Q$, beyond reach in practice. Hence, we chose to compare the subgraphs of $G$ having the same number of nodes as $Q$. We adopted the same setting as [32] for TALE here. For MCS, a subgraph $G_s(V_s, E_s)$ of $G$ matches pattern graph $Q$ if $\frac{|\mathsf{mcs}(Q,G_s)|}{\max(|V_q|,|V_s|)} \geq 0.7$, where $|\mathsf{mcs}(Q, G_s)|$ is the number of nodes in the maximum common subgraph $\mathsf{mcs}(Q, G_s)$ of $Q$ and $G_s$ computed via the algorithm of [25].

The experiments were run on a cluster of 30 machines, all with Intel Core i7 860 CPU and 16GB of memory. Each test was repeated over 5 times, and the average is reported here.

**Experimental results**. We next present our findings. In all the experiments, we fixed $l = 200$, and set $\alpha = 1.2$ by default when generating pattern and data graphs.

**Exp-1: Quality of matches**. In the first set of experiments, we evaluated the quality of matches found by strong simulation vs. matches found by subgraph isomorphism and simulation. We measured the quality of matches as follows.

(1) We first designed pattern graphs, and manually checked the quality of matches returned by Match, VF2 and Sim. We find that Match is able to identify sensible matches. We illustrate this with two example pattern graphs.

Two real-life pattern graphs $Q_A$ and $Q_Y$ are shown in Figures 7(a) and 7(b), respectively. Pattern graph $Q_A$ is to find all "Parenting & Families" books in Amazon network data (a) that are co-purchased with both "Children's Books" and "Home & Garden" books; and (b) that are co-purchased with "Health, Mind & Body" books and vice versa.

Pattern graph $Q_Y$ poses a request on YouTube network data searching for all "Entertainment" videos (a) that are related to "Film & Animation" videos and "Music" videos; and further, (b) for each such "Entertainment" video $x$, there is another "sports" video that is related to the "Film & Animation" and "Music" videos to which $x$ is related.

In data graphs $G_A$ and $G_Y$, nodes are books and videos, respectively, labeled with their ids, and they only match the nodes of $Q_A$ and $Q_Y$ with the same geometry shapes, *e.g.*, circles, ellipses, and regular squares and pentagons.

The match results of $Q_A$ and $Q_Y$ are shown in Figures 7(a) and 7(b), respectively. For pattern graph $Q_A$, subgraph $G_A$ is a sensible match found by Match, but it was not found by VF2. Subgraph $G'_A$ is a match found by Sim in which the "Parenting & Families" books are not co-purchased with both "Children's Books" and "Home & Garden" books, among other things, and was successfully filtered by Match and VF2. These tell us that strong simulation is able to identify sensible matches that subgraph isomorphism fails to catch, and moreover, to eliminate excessive matches by graph simulation that do not make sense.

For pattern graph $Q_Y$, subgraph $G_Y$ is a match found by Match, while subgraphs $G_{Y,1}$, $G_{Y,2}$ and $G_{Y,3}$ are three matches found by VF2. This example shows how strong simulation reduces the sizes of matches found by subgraph isomorphism, without loss of information.

(2) To further measure the quality of matches found, we use:

closeness = #matches_subIso / #matches_found,

where #matches_subIso and #matches_found are the total numbers of nodes in matches found by VF2 and those by a comparative algorithm (Sim, Match, VF2, TALE, MCS), respectively. Recall that matches found by VF2 are also matches found by Match and Sim, by Proposition 1. Hence closeness is essentially the ratio of matched nodes found by VF2 to the entire matches found by Sim, Match, VF2, TALE or MCS. Note that for VF2, closeness is always 1.

(i) To evaluate the impact of pattern graphs $Q$, we fixed $|V|$, *e.g.*, Amazon with 31245 nodes, YouTube with 9368 nodes, and synthetic data with $5 \times 10^4$ nodes, respectively, while varying $|V_q|$ from 2 to 20.

(ii) To evaluate the impact of data graphs $G$, we fixed pattern graphs $Q$ with $|V_q| = 10$ and varied the size of data graphs. We varied $|V|$ from $3 \times 10^3$ to $3 \times 10^4$ nodes for Amazon and from $10^3$ to $10^4$ for YouTube. For synthetic data, we varied $|V|$ from $10^4$ to $10^5$. We used relatively smaller data graphs since VF2 does not scale to large graphs.

The closeness results are reported in Figures 7(c), 7(d), 7(e), 7(f), 7(g) and 7(h). We can see that the closeness of Match is consistently in the range of [70%, 80%] with various query and data graphs, while Sim is in [25%, 38%], TALE is in [35%, 42%], and MCS is in [46%, 57%], respectively. Hence, Match does much better than Sim (up to 50%), TALE (up to 36%) and MCS (up to 23%) at preserving graph topology. Indeed, 70% to 80% of the matches found by Match are exactly those found by VF2, which enforces strict topological matching. Recall that Match is able to find sensible matches missed by VF2 (Examples 1 and 2 and the above quality test (1)). That is, the [20%, 30%] matches found by Match, but missed by VF2, further contain sensible matches.

In addition, the results tell us that the match quality of Match, Sim, TALE and MCS are *not* sensitive to the size of pattern and data graphs for both real-life and synthetic data, a desirable property when match quality is concerned.

(3) In the same setting as (2) above for testing closeness, we tested the numbers of the matched subgraphs in data graphs returned by Match, VF2, TALE and MCS. We did not report Sim since it always returns at most one matched subgraph.

---

[1] http://snap.stanford.edu/data/index.html
[2] http://netsg.cs.sfu.ca/youtubedata/
[3] http://projects.skewed.de/graph-tool/

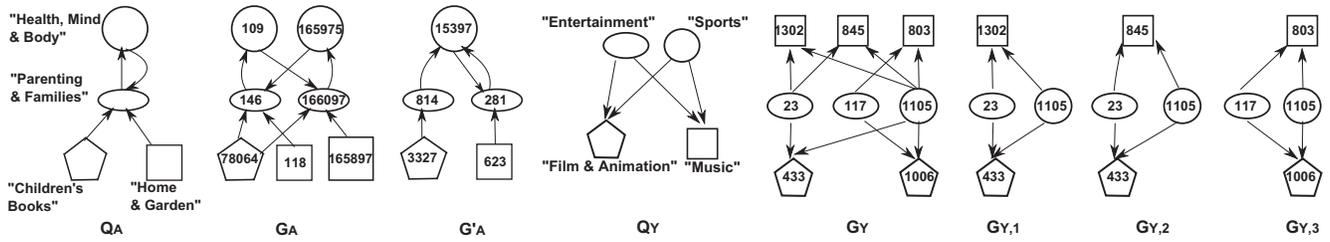

(a) Real-life matches on Amazon  (b) Real-life matches on YouTube

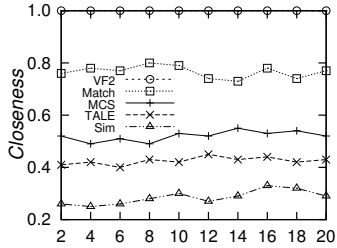 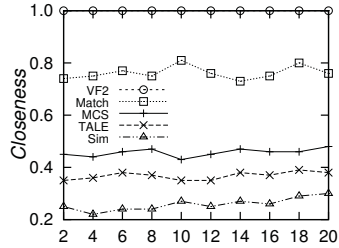 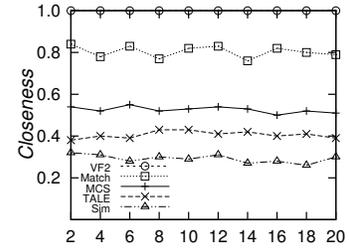

(c) Vary $|V_q|$ (Amazon)  (d) Vary $|V_q|$ (YouTube)  (e) Vary $|V_q|$ (synthetic)

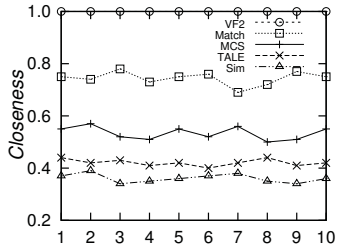 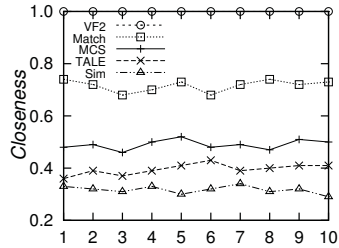 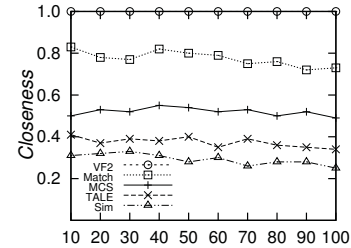

(f) Vary $|V| \times 3 \times 10^3$ (Amazon)  (g) Vary $|V| \times 10^3$ (YouTube)  (h) Vary $|V| \times 10^3$ (synthetic)

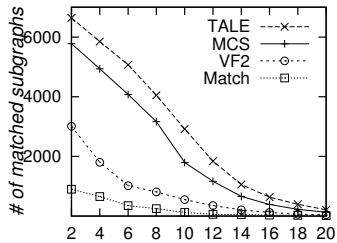 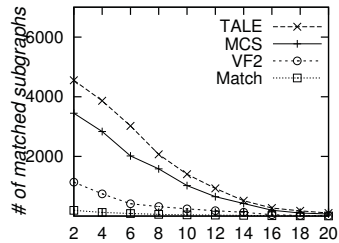 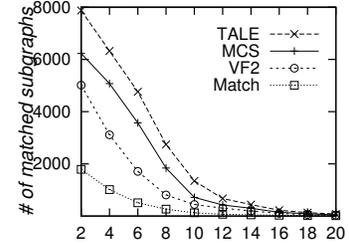

(i) Vary $|V_q|$ (Amazon)  (j) Vary $|V_q|$ (YouTube)  (k) Vary $|V_q|$ (synthetic)

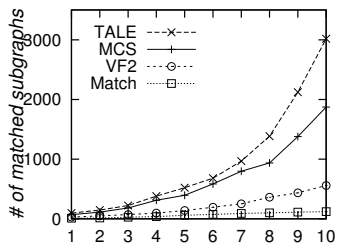 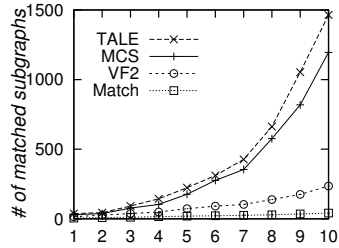 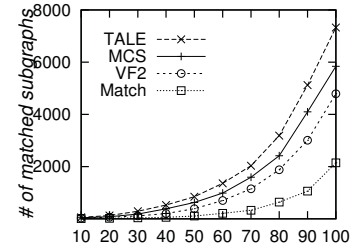

(l) Vary $|V| \times 3 \times 10^3$ (Amazon)  (m) Vary $|V| \times 10^3$ (YouTube)  (n) Vary $|V| \times 10^3$ (synthetic)

**Figure 7: Match quality evaluation on real-life data**

The results are reported in Figures 7(i), 7(j), 7(k), 7(l), 7(m) and 7(n). They tell us that Match returns much less matched subgraphs than VF2: it returns consistently around 25% to 38% matched subgraphs of VF2, for synthetic graph, Amazon and YouTube alike. For approximate matching algorithm TALE and MCS, it is obvious that they return even much more subgraphs than VF2. Indeed, as shown in Fig. 7(n), for example, Match returns 2144 matched subgraphs compared to 4792 by VF2, 5843 by MCS and 7328 by TALE, on a synthetic data graph with $10^5$ nodes. This confirms that Match effectively reduces the sizes of match results, and hence, allows users to effectively analyze the match results on large graphs in practice.

In addition, the number of matched subgraphs decreases when the size of pattern graphs increases, and it increases when the size of data graphs increases, as expected. We also find that although VF2 may find exponentially matches in theory, it does not happen very often in practice.

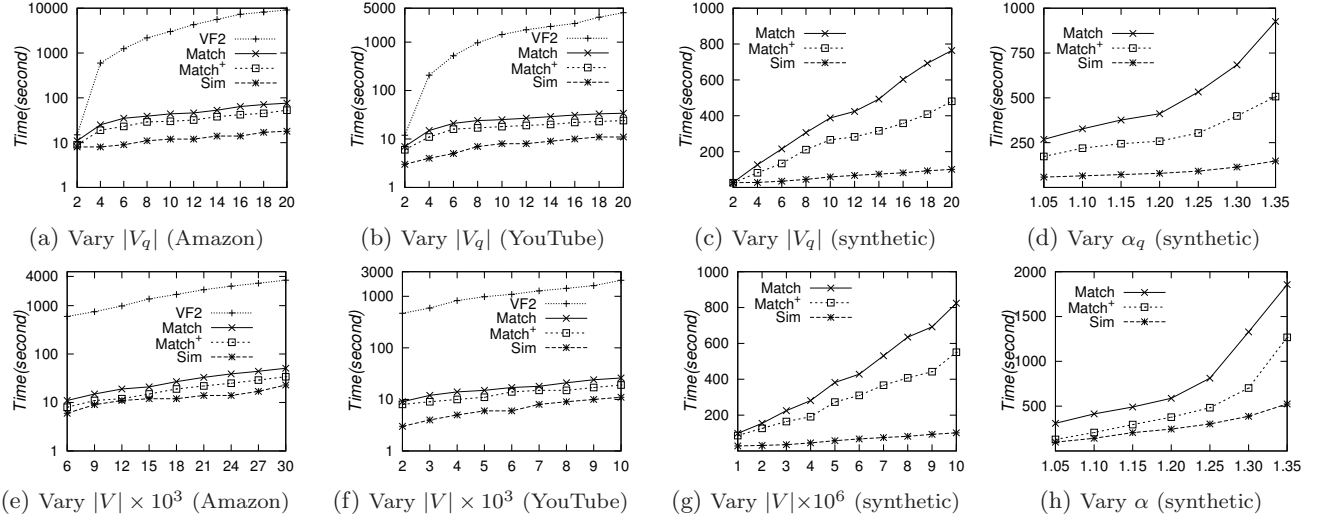

Figure 8: Performance evaluation of centralized algorithms

| #nodes | [0, 9] | [10, 19] | [20, 29] | [30, 39] | [40, 49] | ≥ 50 |
|---|---|---|---|---|---|---|
| Amazon | 0 | 98 | 23 | 0 | 0 | 0 |
| YouTube | 0 | 21 | 18 | 1 | 1 | 0 |
| Synthetic | 0 | 187 | 113 | 65 | 6 | 0 |

Table 3: Sizes of matched subgraphs

(4) In the same setting as (2) for testing closeness with largest possible datasets, e.g., Amazon with 31245 nodes, YouTube with 9368 nodes, and synthetic data with 100000 nodes, we tested the sizes of the matched subgraphs in data graphs returned by Match and Sim.

For Sim, it returns a single matched subgraph with 103, 177 and 311 nodes in Amazon, YouTube and synthetic data, respectively. For Match, the results are reported in Table 3. Their matched subgraphs are typically small, where (a) all matched subgraphs have less than 50 nodes, and (b) over 80% of matches have less than 30 nodes, on real-life and synthetic data. This tells us that strong simulation indeed restricts the sizes of matches, due to the duality and locality.

**Exp-2: Performance of centralized algorithms**. In the second set of experiments, we evaluated the performance of our algorithms Match, Match$^+$ and algorithms Sim and VF2. Algorithm VF2 does not scale well with large data graphs, e.g., it took VF2 more than *three* hours on data graphs with $5 \times 10^6$ nodes (when $\alpha = 1.2$). Hence, we only report the performance of VF2 on the small real-life datasets of Amazon and YouTube that were used to evaluate the quality of matches. For large synthetic data graphs, we only report the other three algorithms Match, Match$^+$ and Sim. In all of our experiments, we also found that TALE and MCS were even much slower than VF2, and hence we did not report the running time of TALE and MCS here.

(1) To evaluate the impact of pattern graphs $Q$, we used two small real-life datasets (Amazon and YouTube) and one large synthetic dataset. We fixed Amazon, YouTube and the synthetic data to have $3 \times 10^4$ nodes, $10^4$ nodes and $5 \times 10^6$ nodes, respectively, while varying the number $|V_q|$ of query nodes from 2 to 20 or the density $\alpha_q$ of pattern graphs from 1.05 to 1.35 (*i.e.,* increasing pattern edges). The results are reported in Figures 8(a), 8(b), 8(c) and 8(d).

The elapsed time of algorithms over real-life datasets is shown in Figures 8(a) and 8(b). When varying $|V_q|$, VF2 is consistently much slower than the other three algorithms in both cases. It is about 100 times slower than Match$^+$ when $V_q \geq 4$ on the two real-life datasets. For instance, it took VF2 hours on the small Amazon and YouTube datasets. Note that, however, when $|V_q| = 2$, VF2 is almost as efficient as the other algorithms. This is consistent with the complexity analysis of VF2: VF2 is in low PTIME when $|V_q| = 2$.

As shown in Fig. 8(c), all these algorithms scale well with $|V_q|$ on large data graphs, except VF2. When we increased the density $\alpha_q$ of pattern graphs, Figure 8(d) shows that these algorithms scale well with the density $\alpha_q$ on large data graphs, except VF2. Algorithms Match and Match$^+$ are slower than Sim, as expected. Indeed, this is a price that has to be paid in exchange for better match quality. We did not report the performance of VF2 in Figures 8(c) and 8(d) since it could not run to completion when $|V_q| \geq 4$.

Finally, observe that the running time of all algorithms increases when $|V_q|$ or $\alpha_q$ increases. This is consistent with the complexity analyses of these algorithms.

(2) To evaluate the impact of data graphs $G$, we also used two small real-life datasets (Amazon and YouTube) and one large synthetic dataset. We fixed pattern graphs with $|V_q| = 10$, while varying the number $|V|$ of nodes of Amazon, YouTube and the synthetic data from $6 \times 10^3$ to $3 \times 10^4$, $2 \times 10^3$ to $10^4$ and $10^6$ to $10^7$, respectively, or varying the density $\alpha$ of data graphs from 1.05 to 1.35. The results are shown in Figures 8(e), 8(f), 8(g) and 8(h).

These results are consistent with the results of varying pattern graph sizes. (a) As shown in Figures 8(e), 8(f), 8(g) and 8(h), all these algorithms except VF2 scale well with the size of data graphs and with the density $\alpha$ of data graphs; (b) algorithms Match and Match$^+$ are slower than Sim; and (c) the running time of VF2 increases far more substantially with the size and density of data graphs than the others. For example, the running time of Match$^+$ increased from about 100s to 600s when the number of nodes of the synthetic data varied from $10^6$ to $10^7$; in contrast, VF2 spent nearly 4000s on Amazon data with $3 \times 10^4$ nodes, but only around 30s on Amazon graphs with $3 \times 10^3$ nodes.

(3) The experimental results in (1) and (2) above also verify that our optimization techniques are effective. Indeed, the

running time of Match$^+$ is consistently about 2/3 of the time taken by Match, a significant reduction.

**Summary**. From these experimental results we find the following. (1) Strong simulation is able to identify sensible matches that are not found by subgraph isomorphism, and eliminate those found by graph simulation but are not meaningful. In addition, it finds high quality matches that retain graph topology. Indeed, 70%-80% of matches found by subgraph isomorphism are retrieved by strong simulation, (up to 50%) better than graph simulation, without paying the price of intractable complexity and large number (or size) of matches. (2) Our algorithms for strong simulation are efficient and scale well with the size and density large-scale data graphs, e.g., it took 270 seconds when $|V| = 10^8$, $|V_q| = 10$ and $|M| = 30$. (3) Our optimization techniques are effective, reducing the running time by at least 33%.

## 6. Conclusion

We have proposed strong simulation to rectify problems of graph pattern matching based on subgraph isomorphism and graph simulation. We have verified, both analytically and experimentally, that strong simulation has several salient features, notably (1) it is capable of capturing the topological structures of pattern and data graphs; (2) it retains the same cubic-time complexity of previous extensions of graph simulation, (3) it demonstrates data locality and allows efficient distributed evaluation algorithms, and (4) it finds bounded matches. Our experimental results have also verified the effectiveness of our optimization techniques.

Several topics are targeted for future work. First, we are to extend strong simulation by incorporating regular expressions on edge types, along the same lines as [18]. Second, our distributed algorithms just aim to demonstrate the data locality of strong simulation. Sophisticated algorithms can be developed in the distributed setting, with better performance guarantees. Third, we are to find metrics to rank matches found by strong simulation, to return top-ranked matches only. Finally, for large graphs, cubic time is still too expensive. We are to explore indexing techniques to speed up the computation, and incremental methods for strong simulation, minimizing unnecessary recomputation in response to (frequent) changes to real-life graphs.

**Acknowledgments**. Fan is supported in part by the RSE-NSFC Joint Project Scheme, an IBM scalable data analytics for a smarter planet innovation award, the National Basic Research Program of China (973 Program) 2012CB316200, and NSFC 61133002. Shuai is supported in part by 973 grant 2011CB302602 and NSFC grants 61170294 and 60903149.